\documentclass[manuscript,natbib=false]{acmart}

\AtBeginDocument{%
  \providecommand\BibTeX{{%
    \normalfont B\kern-0.5em{\scshape i\kern-0.25em b}\kern-0.8em\TeX}}}

\acmConference[ACM Transactions on Quantum Computing]{Make sure to enter the correct
  conference title from your rights confirmation emai}
\acmPrice{15.00}
\acmISBN{978-1-4503-XXXX-X/18/06}

\usepackage{algorithm}
\usepackage{algpseudocode}
\usepackage[T1]{fontenc}
\usepackage{subcaption}
\usepackage{adjustbox}

\usepackage{tikz}
    \usetikzlibrary{quantikz}
    \usetikzlibrary{shapes.geometric, arrows, shadows}
\tikzstyle{B} = [decorate, decoration={brace,amplitude=4mm, raise=2mm,##1, mirror} ]

\tikzstyle{startstop} = [rectangle, rounded corners, minimum width=2cm, minimum height=0.7cm,text centered, draw=black, fill=red!30]
\tikzstyle{io} = [trapezium, trapezium left angle=70, trapezium right angle=110, minimum width=2cm, minimum height=0.7cm, text centered, draw=black, fill=blue!30]
\tikzstyle{process} = [rectangle, minimum width=2cm, minimum height=0.7cm, text centered, draw=black, fill=orange!30]
\tikzstyle{decision} = [diamond, minimum width=2cm, minimum height=0.7cm, text centered, draw=black, fill=green!30]
\tikzstyle{arrow} = [thick,->,>=stealth]

\definecolor{codegreen}{rgb}{0,0.6,0}
\definecolor{codegray}{rgb}{0.5,0.5,0.5}
\definecolor{codepurple}{rgb}{0.58,0,0.82}
\definecolor{backcolour}{rgb}{0.95,0.95,0.92}

\definecolor{listinggray}{gray}{0.9}
\definecolor{lbcolor}{rgb}{0.9,0.9,0.9}
\definecolor{Darkgreen}{rgb}{0,0.4,0}

\RequirePackage[
datamodel=acmdatamodel,
style=acmauthoryear,
sorting=none
]{biblatex}
\begin{document}

\title[Classical programming principles in the Intel Quantum SDK]{Utilizing classical programming principles in the Intel Quantum SDK: implementation of quantum lattice Boltzmann method}

\author{Tejas Shinde}
\email{tejas.shinde@quanscient.com}
\affiliation{%
  \institution{Quanscient Oy}
  \streetaddress{Åkerlundinkatu 8}
  \city{Tampere}
  \state{Pirkanmaa}
  \country{Finland}
  \postcode{33100}
}
\affiliation{%
  \institution{University of Jyväskylä}
  \streetaddress{Mattilanniemi 2}
  \city{Jyväskylä}
  \state{Keski-Suomi}
  \country{Finland}
  \postcode{40100}
}
\author{Ljubomir Budinski}

\affiliation{%
  \institution{Quanscient Oy}
  \streetaddress{Åkerlundinkatu 8}
  \city{Tampere}
  \state{Pirkanmaa}
  \country{Finland}
  \postcode{33100}
}
\affiliation{%
  \institution{Faculty of Technical Sciences, University of Novi Sad}
  \streetaddress{Trg Dositeja Obradovića 6}
  \city{Novi Sad}
  \country{Serbia}
  \postcode{21000}
}
\email{ljubomir.budinski@quanscient.com}

\author{Ossi Niemimäki}

\affiliation{%
  \institution{Quanscient Oy}
  \streetaddress{Åkerlundinkatu 8}
  \city{Tampere}
  \state{Pirkanmaa}
  \country{Finland}
  \postcode{33100}
}
\email{ossi.niemimaki@quanscient.com}
\author{Valtteri Lahtinen}

\email{valtteri.lahtinen@quanscient.com}
\affiliation{%
  \institution{Quanscient Oy}
  \streetaddress{Åkerlundinkatu 8}
  \city{Tampere}
  \state{Pirkanmaa}
  \country{Finland}
  \postcode{33100}
}

\author{Helena Liebelt}
\affiliation{%
  \institution{Technische Hochschule Deggendorf}
  \streetaddress{Dieter-Görlitz-Platz 1}
  \city{Deggendorf}
  \country{Germany}}
\email{helena.liebelt@th-deg.de}
\author{Rui Li}
\email{rui.li@th-deg.de}
\affiliation{%
  \institution{Technische Hochschule Deggendorf}
  \streetaddress{Dieter-Görlitz-Platz 1}
  \city{Deggendorf}
  \country{Germany}}


\begin{abstract}
  We explore the use of classical programming techniques in implementing the quantum lattice Boltzmann method in the Intel Quantum SDK -- a software tool for quantum circuit creation and execution on Intel quantum hardware. 
  As hardware access is limited, we use the state vector simulator provided by the SDK.
  The novelty of this work lies in leveraging classical techniques for the implementation of quantum algorithms. 
  We emphasize the refinement of algorithm implementation and devise strategies to enhance quantum circuits for better control over problem variables. 
  To this end, we adopt classical principles such as modularization, which allows for systematic and controlled execution of complex algorithms. Furthermore, we discuss how the same implementation could be expanded from state vector simulations to execution on quantum hardware with minor adjustments in these configurations.
   
\end{abstract}




\maketitle

\section{Introduction}
In the field of engineering, industrial and scientific simulations place substantial demands on computational resources. Advances in computing hardware, such as high-performance computing clusters and GPUs, are catering to increasing computational demands, but the ability to shrink transistors is hitting physical limits, threatening the well-known Moore’s law~\cite{moorelaw}. Quantum computing has received substantial attention over the past few years as a means for pushing the envelope. New algorithms have been developed, showing algorithmic efficiency over classical computing. For example, there is a potential
quantum speed-up in applications such as linear algebra~\cite{lloyd2020quantum,qian2019quantum,ambainis2010variable}, quantum chemistry~\cite{Cao_2019}, optimization~\cite{Kerenidis_2020,farhi2014quantum} and machine learning~\cite{garg2020advances,sharma2020qeml}.

With the development of algorithms, there is a need for refined methods and techniques for smooth and robust implementation. Implementing the quantum circuits in a reusable and scalable manner, and integrating them into possible applications poses a new challenge. At its core, this challenge arises from the intricate nature of quantum algorithms, which usually rely on complex sequences of quantum gates acting on qubits. Achieving reusability is challenging due to the constraints and the specificity of each application. Scalability is vital for building larger optimized quantum algorithms and ensuring efficient utilization of the resources as systems grow in size.

The main contribution of this work is the use of classical programming principles in implementing quantum algorithms in the Intel Quantum SDK~\cite{IntelSDK}. The Intel Quantum SDK is a C\texttt{++} based quantum programming framework tailored for hybrid quantum-classical programming. For this work, we have chosen the quantum lattice Boltzmann method (QLBM) as an example for the implementation. The QLBM is a promising quantum algorithm for computational fluid dynamics simulations. It provides a quantum-native way to solve transport phenomena, with recent advancements showing its potential for quantum efficiency \cite{ADE, NavierQLBM, budinski2023efficient}. In this paper, we focus on the QLBM algorithm for the advection-diffusion equation as presented in \cite{ADE}. 
Additionally, we provide a brief explanation on expanding the implementation to 2D advection-diffusion~\cite[Section~3.2]{ADE} and Navier-Stokes equations using the stream function-vorticity formulation~\cite{NavierQLBM}. However, we do not aim to expand or improve the QLBM algorithm itself in this work, but rely solely on the already-existing knowledge.

Through this example, we explore the programming design principles in the context of quantum algorithms. We focus on the modular approach in forming generic building blocks for the complex algorithm, such that can be easily scaled and reused in different configurations. We also outline possibilities for hybrid design, for which the Intel Quantum SDK is particularly suitable.

This paper is structured as follows. Section~\ref{sec:Intel} provides a concise overview of the Intel Quantum SDK. In Section~\ref{sec:qlbm} we briefly recap the lattice Boltzmann method and discuss the quantum circuit implementation in general. In Section~\ref{sec:algorithm}, we begin with an analysis of how classical techniques are employed for the implementation of quantum algorithms in the Intel Quantum SDK. The section concludes with an exploration of expanding the implementation to address 2D advection-diffusion and Navier-Stokes equations, accompanied by numerical validation using a simple example.

\section{Intel Quantum SDK}
\label{sec:Intel}

The Intel Quantum SDK is an LLVM\footnote{A collection of compiler and toolchain technologies.} based environment that allows the writing of hybrid quantum-classical algorithms on computational systems, targeting variational quantum algorithms in particular~\cite{IntelSDK}. The SDK follows a similar approach as a classical hardware accelerator environment, which allows one to write instructions targeted for specialized hardware such as GPU and FPGA in order to increase efficiency.
The SDK aims to make it efficient to run hybrid quantum-classical parts of computational problems, where the quantum hardware acts as an accelerator. 
The Intel Quantum SDK enables quantum kernels for quantum instructions targeted for the quantum hardware, and classical methods for the classical instructions targeted towards the CPU within the same algorithm.
Separating the instructions benefits the use of hybrid quantum and classical computing.

\begin{figure}[ht]
    \centering
    
\begin{tikzpicture}[
    node distance=0.05cm,
    every join/.style={->, thick},
    stage1/.style={rectangle, draw=white, rounded corners, minimum width=8cm, minimum height=1cm, align=center, fill=blue!20!black!80!white},
    stage2/.style={rectangle, draw=white, rounded corners, minimum width=8cm, minimum height=1cm, align=center,fill=blue!50!black!50!white},
    stage3/.style={rectangle, draw=white, rounded corners, minimum width=8cm, minimum height=1cm, align=center, fill=blue!35!black!65!white, text width=2cm},
    stage4/.style={rectangle, draw=white, rounded corners, minimum width=5.33cm, minimum height=1cm, align=center, fill=blue!50!black!50!white},
    stage5/.style={rectangle, draw=white, rounded corners, minimum width=5.33cm, minimum height=1cm, align=center, fill=blue!50!black!50!white, text width=2cm},
    process/.style={rectangle, draw=white, rounded corners, minimum width=2.6cm, minimum height=1cm, align=center, fill=blue!60!black!40!white, text width=2.4cm},
    process1/.style={rectangle, draw=white, rounded corners, minimum width=2.6cm, minimum height=3.125cm, align=center, fill=blue!60!black!40!white, text width=2.4cm},
    block/.style = {rectangle, dotted, draw, rounded corners, fill=yellow!20, fill opacity = 0.7, text width=3cm, minimum height=25mm, minimum width=\textwidth, align=flush left},
]
\newcommand{\textboldcolor}[2]{\textbf{\textcolor{#1}{#2}}}
\centering

\node[stage1] (start) {\textboldcolor{white}{Application} \\ \textboldcolor{white}{\small{\textit{hybrid quantum-classical unified C++ source file}}}};
\node[stage3, , below=of start] (step2) {\textboldcolor{white}{Compiler}};
\node[stage2, , below=of step2] (step3) {\textboldcolor{white}{Quantum Runtime}};
\node[stage4, , below=of step3, xshift=1.33cm] (step4) {\textboldcolor{white}{Qubit control Processor}};
\node[stage5, , below=of step4] (step5) {\textboldcolor{white}{Control Electronics}};
\node[process, , below=of step3, yshift=-2.125cm] (step6) {\textboldcolor{white}{Quantum Dot Simulator}};
\node[process, , below=of step3, yshift=-2.125cm, xshift=2.66cm] (step7) {\textboldcolor{white}{Quantum Dot Chip}};
\node[process1, , below=of step3, xshift=-2.66cm] (step8) {\textboldcolor{white}{Intel Quantum Simulator (IQS)}};
\end{tikzpicture}
\caption{Overview of Intel Quantum SDK as described in \protect{\cite{IntelSDK}}.}
\Description{A block diagram illustrating the layers of a hybrid quantum-classical computing stack. The top layer is labeled "Application" and is described as a hybrid quantum-classical unified C++ source file. Below it is the "Compiler" layer. Underneath the compiler is the "Quantum Runtime" layer. The next layer is the "Qubit Control Processor," followed by the "Control Electronics" layer. The bottom layer is divided into three components: "Intel Quantum Simulator (IQS)," "Quantum Dot Simulator," and "Quantum Dot Qubit Chip." The diagram shows the flow from the application layer down to the hardware or simulation components.}
\label{fig:OverviewIntelSDK}
\end{figure}
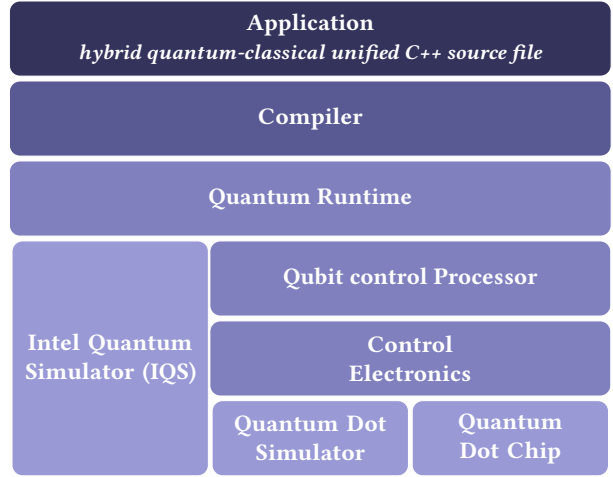

The SDK provides a C\texttt{++} based programming interface based on the circuit model for quantum computation. The SDK is developed with the Intel quantum hardware in mind, but these devices are not yet available to general public. Instead, the SDK provides a full-state simulator class with API calls to set up a CPU-based simulator and provides access the quantum state during the simulation. It allows users to simulate and test their program in an ideal case without access to actual quantum hardware. The quantum state is accessed through methods that return the conditional probabilities of the qubits and complex amplitudes of the state space. Fig.~\ref{fig:OverviewIntelSDK} gives an overview of the Intel quantum computing stack.

The version 1.0 of the Intel Quantum SDK added a Python interface to the existing framework. This helps convert OpenQASM2~\cite{cross2017open} circuit instructions to Intel Quantum SDK compatible C\texttt{++} source code. The Python interface also allows to compile and run the quantum circuits directly. We make use of this in the implementation to add circuit functionalities that would have been otherwise difficult to realize through the C\texttt{++} interface at the time of writing.

\section{Quantum lattice Boltzmann method}
\label{sec:qlbm}

The lattice Boltzmann method (LBM) is a computational fluid dynamics technique for simulating complex fluid flows. Traditional methods in computational fluid dynamics, such as finite difference, finite volume, and finite element methods, rely on solving the Navier-Stokes equations to model fluid flow. These techniques involve complex discretization schemes and extensive computational resources, often leading to challenges in handling complex boundary conditions and multiphase flows. In contrast, the LBM offers a more intuitive and flexible approach by simulating fluid dynamics through the evolution of particle distribution functions on a lattice grid -- see for example \cite{AAM} for a comparison of the LBM with the traditional finite difference method in different scenarios. 
 
In particular, we are interested in the LBM as a promising method for solving fluid dynamics problems with quantum computers, and use the simple advection-diffusion model~\cite{ADE} as an example of a hybrid computation scheme. For more recent advancements on quantum implementations see for example \cite{Collisionless-QLBM} and ~\cite{Schalkers:2024jtp}.

The LBM operates on a lattice grid where each node represents a fluid particle distribution function. Through a series of collision and propagation steps, these functions evolve over time to simulate fluid behavior. 
 
The LBM follows a mesoscopic approach and bridges the gap between molecular dynamics and macroscopic fluid dynamics, offering insights into various flow phenomena at different scales.~\cite{THELBM}

\begin{figure}[h]
\centering
\begin{tikzpicture}
    
    \draw (1,0.2) -- (1,-0.2);
    \fill[black] (0,0) circle(.12);
    \fill[black] (2,0) circle(.12);
    \draw[arrows = {-Stealth[length=6pt, inset=1.2pt]}] (1,0)-- (0.1,0) ;
    \draw[arrows = {-Stealth[length=6pt, inset=1.2pt]}] (1,0) -- (1.9,0);
    
    \node at (1,-0.5) {D1Q2};
    \node at (0.5,0.5) {2};
    \node at (1.5,0.5) {1};
\end{tikzpicture}
\caption{1D lattice arrangement}
\label{latticearrangement}
\Description{The diagram represents a D1Q2 lattice used in the lattice Boltzmann method. This model consists of one-dimensional lattice nodes where each node (here node depicted by the vertical line) has two possible directions for particle movement. The diagram has two velocities labeled "1" and "2." An arrow points from the node to the neighbouring nodes (depicted by circles), indicating the direction of particle movement.}
\end{figure}
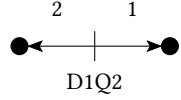

A D1Q2 lattice arrangement is shown in Fig.~\ref{latticearrangement}, where DnQm stands for $n$ dimensions and $m$ velocity vectors.

The D1Q2 arrangement consists of two velocity vectors $e_1 = 1, e_2 = -1$ for the distributions $f_1$ and $f_2$, respectively. At an instance in time, two particles exist at a site, moving right and left in the streaming process. The weighting factors for the D1Q2 are $1/2$ for both distributions $f_i$.

As a simple example, we study the QLBM solution to the advection-diffusion equation. The quantum algorithm for the D1Q2 QLBM can be divided into four major steps: encoding, collision, propagation, and the calculation of macroscopic variables. Each of these is illustrated in the circuit implementation in Fig.~\ref{fig:Fullcircuit}. The steps proceed as follows:
\begin{enumerate}{\label{steps}}
\item Encoding: 
    In the encoding step we set up two quantum registers $q$ and $a$. To encode the vector into quantum states and do the computation we require the quantum working register $q$. The working register encodes the vector with concentration $C(x,t)$ having qubits equal to $\log_{2}(2M)$, where $M$ is the lattice size. The ancilla register $a$ consists of only one qubit and is used in the controlled operation in the collision step, as shown in Fig.~\ref{fig:Fullcircuit}. To achieve the initial state, external state preparation is used, for example the state preparation procedure proposed by Shende et al.~\cite{Shende}
\item Collision:
    The collision step is a point-wise multiplication between the vector $C$ and the block diagonal matrix according to 
    \begin{equation}\label{feqade}
     f_i^{eq}(\vec{x},t) = \omega_i C(x,t) \left( 1+ \frac{e_i \cdot \vec{u}}{c_s^{2}} \right) ,
    \end{equation}
    excluding the depended variable vector $C(x,t)$. Since the block diagonal matrix is non-unitary, linear combination of unitaries approach is used, providing two unitary diagonal operators $A_1$ and $A_2$. The quantum circuit for the collision part is given in Fig.~\ref{fig:Fullcircuit} in between the states $|\psi_0 \rangle$ and $|\psi_1 \rangle$. The decomposition of the $A_1$ and $A_2$ gates in the circuit Fig.~\ref{fig:Fullcircuit} is given in Fig.~\ref{operator} involving a phase gate and an $X$ gate. 
\item Propagation:
    In the propagation step we use the quantum walk~\cite{QW,Low2019hamiltonian} procedure for the streaming of the distributions. In this step the distribution function $f_1$ moves to the right with speed $e_1 = 1$ and the distribution function $f_2$ moves to the left with speed $e_2 = -1$. This is achieved by the operators illustrated in the Fig.~\ref{fig:Fullcircuit}, $R$ and $L$  for right shift and left shift respectively.
\item Macroscopic variables:
    The last step involves the calculation of the macroscopic variables, density or velocity, this is possible through the point-wise addition of the two quantum states. As the equation~\cite[Eq.~11]{ADE} indicates that both of the distribution functions are located in the subsystem controlled by the ancilla qubit $|0\rangle_a$, for the point-wise addition at first a swap gate is added between the last qubit $q_n$ of the working register $q$ and the ancilla qubit $a$. This operation switches the states of the working register controlled by the $|0\rangle_a$ and $|1\rangle_a$ in the ancilla register, resulting in the positioning of the sub-state of the second distribution function, controlled by $|0\rangle_a$, to the state controlled by $|1\rangle_a$. The implementation of this is simple as given in the Fig.~\ref{fig:Fullcircuit}, just the swap and a Hadamard gate on the ancilla at the end.
\end{enumerate}
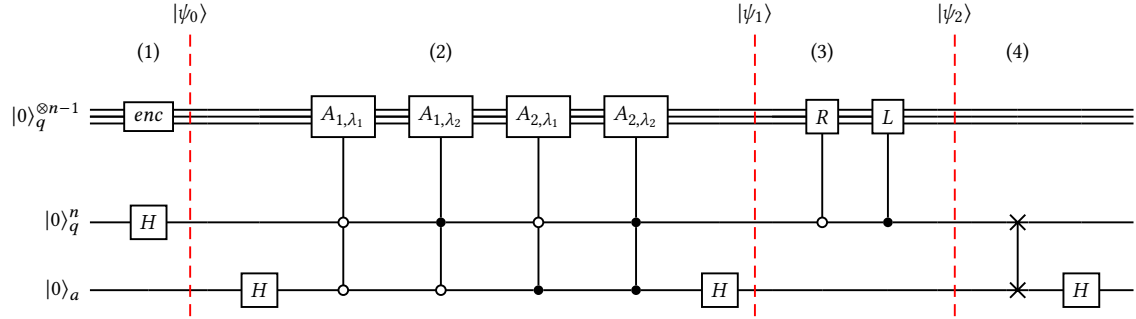
\begin{figure}[!ht]
\centering
\begin{adjustbox}{width=1.0\textwidth}
\begin{quantikz}
    &{{\text{(1)}}} &&&&{{\text{(2)}}} &&&&& {{\text{(3)}}} & &&& {{\text{(4)}}}\\
    \lstick{$\ket{0}_q^{\otimes n-1}$} & \gate{\textrm{${e n c}$}} \qwbundle[alternate]{} &  \qwbundle[alternate]{} & \qwbundle[alternate]{} & \gate{\textrm{$A_{1,\lambda_1}$}} \qwbundle[alternate]{} & \gate{\textrm{$A_{1,\lambda_2}$}} \qwbundle[alternate]{} & \gate{\textrm{$A_{2,\lambda_1}$}} \qwbundle[alternate]{} & \gate{\textrm{$A_{2,\lambda_2}$}} \qwbundle[alternate]{}  &  \qwbundle[alternate]{} & \qwbundle[alternate]{} & \gate{\textrm{$R$}} \qwbundle[alternate]{} &  \gate{\textrm{$L$}} \qwbundle[alternate]{} & \qwbundle[alternate]{} & \qwbundle[alternate]{} & \qwbundle[alternate]{} & \qwbundle[alternate]{} & \qwbundle[alternate]{}\\
    \\
    \lstick{$\ket{0}_{q}^{n}$} & \gate{H}\slice{$|\psi_0\rangle$} & \qw & \qw & \octrl{-2} & \ctrl{-2} & \octrl{-2} & \ctrl{-2} & \qw & \qw & \octrl{-2} & \ctrl{-2} & \qw\slice{$|\psi_2\rangle$} & \qw & \swap{0}  & \qw & \qw \\
    \lstick{$\ket{0}_a$} & \qw & \qw & \gate{H} & \octrl{-1} & \octrl{-1} & \ctrl{-1} & \ctrl{-1} & \gate{H}\slice{$|\psi_1\rangle$} & \qw & \qw & \qw & \qw & \qw & \swap{-1} & \gate{H} & \qw
\end{quantikz}
\end{adjustbox}
    \caption{Quantum circuit to solve the 1D advection-diffusion equation with the lattice Boltzmann method.}
    \label{fig:Fullcircuit}
    \Description{.}
\end{figure}
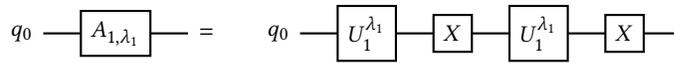
\begin{figure}[!ht]
    \centering
    \begin{quantikz}
       \lstick{$q_0$} & \gate{\textrm{$A_{1,\lambda_1}$}} & \push{\rule{.2em}{0em}=\rule{.2em}{0em}} & &\lstick{$q_0$} & \gate{\textrm{$U_1^{\lambda_1}$}} & \gate{\textrm{$X$}} & \gate{\textrm{$U_1^{\lambda_1}$}} & \gate{\textrm{$X$}} &  \qw 
    \end{quantikz}
    \caption{Quantum circuit to implement the $A_{1,\lambda_1}$ operator.}
    \label{operator}
    \Description{The figure show a generalized quantum circuit to solve the 1D Quantum Lattice Boltzmann Method. The 1D Quantum Lattice Boltzmann Method (QLBM) advection-diffusion quantum circuit begins with the initial state preparation where qubits are set to |0⟩ and encoded (enc). The encoded qubits then pass through a series of advection-diffusion collision operators. The state is then processed through streaming right (R) and shift (L) operations. Swap and Hadamard gate is applied to the ancillary qubit in the macroscopic calculation step, and the circuit concludes with a measurement step. This sequence effectively simulates the advection-diffusion process using quantum gates.}
\end{figure}

The LBM is a time-marching algorithm, and the above outline covers only one time step. It is not detailed in \cite{ADE} how to connect time steps purely inside a quantum circuit, and we will not attempt to present such a solution here. We use the re-encoding of the initial state at each time step as an example of a hybrid design and study the implementation from this particular angle. Note that following this approach likely destroys the possible speed-up advantages of the quantum algorithm. However, it provides an interesting point of study from the program design perspective, which can be useful even after the time-stepping problem for the QLBM has been resolved. Recent attempts towards efficient time-step concatenation can be found, for example, in \cite{Collisionless-QLBM}.

The gate complexity of the  QLBM depends on the resolution of the computational domain, as detailed in \cite[Section~3.3]{ADE}. It is highlighted that the encoding step is the most resource-intensive, requiring a significant number of state preparation steps and $CX$ gates, which depends on the problem to be solved -- in the most general case, this cost is exponential. In the collision step, the complexity is not directly affected by the resolution of the computational domain: increasing the number of qubits to achieve a higher resolution does not change the size of the ancillary register, and the number of multi-qubit gate operations required in the $q$ register is constant.
The propagation step is influenced by the resolution of the computational domain, and the depth of this subroutine increases logarithmically. The complexity of the propagation step can be optimized for example by using the parallel state shift method~\cite{budinski2023efficient}. The overall complexity of the algorithm, excluding the encoding step, is determined to be $O(\log_2(\alpha D)$, where $\alpha$ represents the number of velocity vectors and $D$ depends on the domain resolution ($D=2M$ for $D1Q2$).

\section{Quantum algorithm implementation in Intel Quantum SDK}
\label{sec:algorithm}

In the form presented above, the quantum lattice Boltzmann algorithm is a hybrid method, although not a variational one. This hybrid property makes it, nevertheless, an interesting example for implementation on platforms that support hybrid classical and quantum computing, such as the Intel Quantum SDK. 
 
In this section, we detail such an implementation and explore the use of modularization with the aim of creating easily reusable and scalable quantum circuits.

\subsection{Implementation}

Due to the complex nature of the quantum lattice Boltzmann algorithm, it becomes essential to employ specific quantum software engineering techniques for the implementation. A general idea is proposed in \cite[Section~3]{iterativecycles} about the iterative development cycle and infrastructure for quantum program development: to construct adaptable quantum circuits optimised for execution. In this section, we focus on cycle 2, especially the implementation, the quantum algorithm design cycle, where we adhere to an iterative process for creating and testing various configurations. A modular design methodology is followed for the implementation in the Intel Quantum SDK, similar to one described in \cite{QSE}. 

\begin{figure}[ht]
\centering
\begin{tikzpicture}[node distance=1.25cm]

\node (start) [startstop] {start};
\node (initdist) [io, below of=start] {no. of iterations};
\node (forloop) [decision, below of=initdist, yshift=-0.25cm] {i less iter};
\node (ifelse) [decision, below of=forloop, yshift=-1.5cm] {if iteration == 0};

\node (initvector) [process, below of=ifelse,xshift=-2cm, yshift=-0.5cm]{Find initial vector};
\node (norm) [process, below of=initvector]{ normalize };
\node (decompose) [process, below of=norm]{export to OpenQASM};
\node (norm2) [process, below of=ifelse, xshift=2cm, yshift=-0.5cm]{ normalize };
\node (decompose2) [process, below of=norm2]{export to OpenQASM};

\node (conversion) [process, below of=decompose, xshift=2cm]{OpenQASM $\rightarrow$ C\texttt{++}};
\node (calc) [process, below of=conversion]{QLBM ADE Intel Quantum SDK};
\node (extract) [io, below of=calc]{extract state vector};
\node (stop) [startstop, left of=forloop, xshift=-1cm] {stop};

\draw [arrow] (start) -- (initdist);
\draw [arrow] (initdist) -- (forloop);
\draw [arrow] (forloop) -- node[anchor=east]{yes}(ifelse);
\draw[arrow] (ifelse) -| node[anchor=east]{yes}(initvector);
\draw[arrow] (ifelse) -| node[anchor=west]{No}(norm2);
\draw[arrow] (initvector) -- (norm);
\draw[arrow] (norm2) -- node[anchor=west]{To Qiskit}(decompose2);
\draw[arrow] (norm) -- node[anchor=west]{To Qiskit}(decompose);
\draw[arrow] (decompose) |- (conversion);
\draw[arrow] (decompose2) |- (conversion);
\draw[arrow] (conversion) -- node[anchor=west]{To Intel Quantum SDK}(calc);
\draw[arrow] (calc) -- (extract);
\draw [arrow] (extract.east) to ++ (1.5,0) |- (forloop);
\draw[arrow] (forloop) -- node[anchor=south]{no}(stop);

\draw[decoration={brace,mirror,amplitude=3mm, raise=1.5mm},decorate]
  (-4,-7) -- node[left=1pt, xshift=-0.5cm] {\Large $Pre-processing$} (-4,-11.5);
\draw[decoration={brace,mirror,amplitude=3mm, raise=1.5mm},decorate]
  (-4,-11.6) -- node[left=6pt, xshift=-0.5cm] {\Large $Solver$} (-4,-12.8);
\draw[decoration={brace,mirror,amplitude=3mm, raise=1.5mm},decorate]
  (-4,-12.9) -- node[left=6pt, xshift=-0.5cm] {\Large $Post processing$} (-4,-14);

\end{tikzpicture}
\caption{Overview on the hybrid time-stepping.}
\label{fig:Timestepping}
\Description{.}
\end{figure}
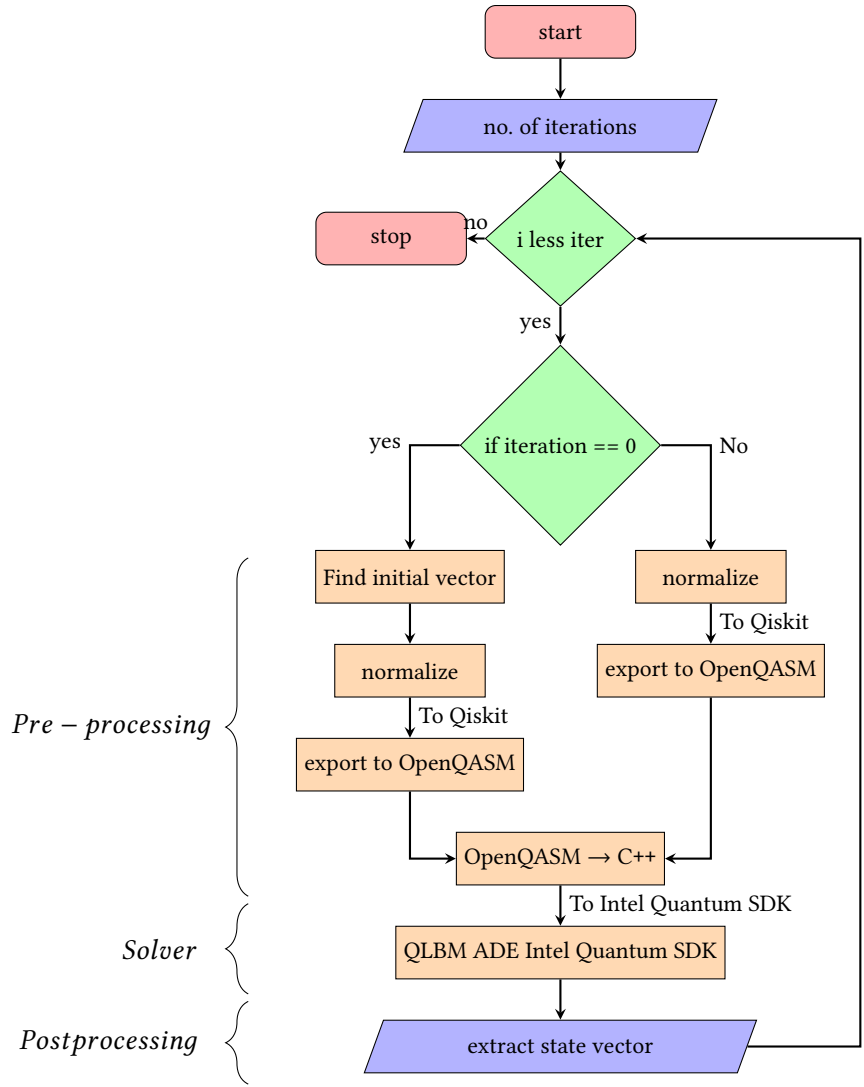

The quantum circuit for the algorithm is provided in Fig.~\ref{fig:Fullcircuit}. Fig.~\ref{fig:Timestepping} illustrates how the hybrid time-stepping is performed in the algorithm. The time-step process explained here is specifically for the purpose of implementing the algorithm in the Intel Quantum SDK, and we have chosen it as an example of the hybrid approach in the algorithm design. To implement the time-stepping, the algorithm is divided into three modules, following the modular design methodology~\cite[Section 5.4]{QSE}, as described below.  
\begin{enumerate}
    \item Pre-processing: preparing the initial encoding circuit.
    \item Solver: solving the four steps discussed in Section~\ref{steps} in the Intel Quantum SDK.
    \item Post-processing: extracting the state vector from the Intel Quantum SDK and post-processing the data.
\end{enumerate}

\subsubsection{The top-down approach}\label{approach}

In this study, we employ the top-down approach to software engineering, adhering to the Single Responsibility Principle. This methodology begins with a high-level design of the overall system, ensuring that each module has a single responsibility. The system is progressively decomposed into smaller, more detailed components, with each module focusing on a specific task. This decomposition promotes modularization, facilitating easier maintenance and scalability.

The Intel Quantum SDK provides a set of quantum logic gates that can be used to create quantum circuits, with each circuit forming a quantum kernel. The goal is to use these quantum kernels and create mutable quantum circuits (adaptable to different configurations) for the different steps involved in the QLBM solver. The hardware accelerator design helps us in the modularization of the quantum circuit in Fig.~\ref{fig:Fullcircuit} and to write a quantum kernel for each module, which we will call \emph{subcircuits}. The different subcircuits are according to the steps involved such as the encoding, collision, propagation and the calculation of macroscopic variables.

The subcircuits might include gates controlled by multiple qubits. All but one of the quantum logic gates available in the Intel Quantum SDK are one- and two-qubit gates, the exception being the three-qubit Toffoli gate. For the implementation of multi-qubit controlled gates, the decomposition method described in the book \cite[Section~4.3]{QuantumBible} is utilized. 
We call such decomposed gates \emph{custom gates}. With the help of the basic logic gates and the custom gates, we have the basic building blocks for the subcircuits, minimising the size of the modules.

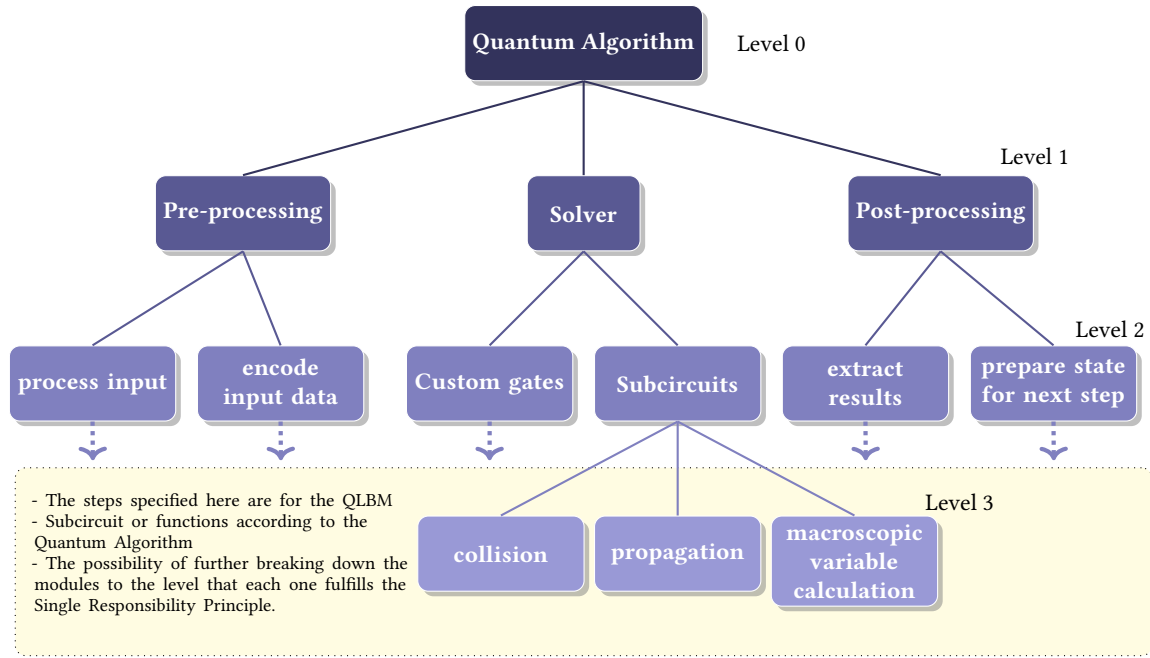
\begin{figure}[ht]
    \centering
    
\begin{tikzpicture}[
    node distance=1.25cm,
    every join/.style={->, thick},
    stage1/.style={rectangle, draw=white, rounded corners, minimum width=1.5cm, minimum height=1cm, align=center, fill=blue!20!black!80!white , drop shadow},
    stage2/.style={rectangle, draw=white, rounded corners, minimum width=1.5cm, minimum height=1cm, align=center, fill=blue!35!black!65!white, drop shadow},
    stage3/.style={rectangle, draw=white, rounded corners, minimum width=2cm, minimum height=1cm, align=center, fill=blue!50!black!50!white, drop shadow, text width=2cm},
    process/.style={rectangle, draw=white, rounded corners, drop shadow, minimum width=2cm, minimum height=1cm, align=center, fill=blue!60!black!40!white, text width=2cm},
    block/.style = {rectangle, dotted, draw, rounded corners, fill=yellow!20, fill opacity = 0.7, text width=3cm, minimum height=25mm, minimum width=\textwidth, align=flush left},
]
\newcommand{\textboldcolor}[2]{\textbf{\textcolor{#1}{#2}}}
\centering
\node[stage1] (start) {\textboldcolor{white}{Quantum Algorithm}};
\node[block=40em, yshift=-6.9cm] (A) {};
\node [text width=5cm] at (-4.8,-6.8) {\footnotesize - The steps specified here are 
for the QLBM - Subcircuit or functions according to the Quantum Algorithm \\
- The possibility of further breaking down the modules to the level that each one fulfills the Single Responsibility Principle.\\ };
\node at (2.5,0) {Level 0};
\node at (6,-1.5) {Level 1};
\node at (7,-3.8) {Level 2};
\node at (5,-6.1) {Level 3};

\node[stage2, below=of start] (solver) {\textbf{\textcolor{white}{Solver}}};
\node[stage2, left=of solver, xshift=-1.35cm] (preprocessing) {\textbf{\textcolor{white}{Pre-processing}}};
\node[stage2, right=of solver, xshift=1.5cm] (postprocessing) {\textbf{\textcolor{white}{Post-processing}}};

\node[stage3, below=of preprocessing, xshift=-2.0cm] (processInput) {\textboldcolor{white}{process input}};
\node[stage3, below=of preprocessing, xshift=0.5cm] (encodeInput) {\textboldcolor{white}{encode input data}};


\node[stage3, below=of solver, xshift=-1.25cm] (customgates) {\textboldcolor{white}{Custom gates}};
\node[stage3, below=of solver, xshift=1.25cm] (subcircuits) {\textboldcolor{white}{Subcircuits}};
\node[process, below=of subcircuits, xshift=-2.35cm] (collision) {\textboldcolor{white}{collision}};
\node[process, below=of subcircuits, ] (propagation) {\textboldcolor{white}{propagation}};
\node[process, below=of subcircuits, xshift=2.35cm] (macroCalc) {\textboldcolor{white}{macroscopic variable calculation}};

\node[stage3, below=of postprocessing, xshift=-1cm] (saveResults) {\textboldcolor{white}{extract results}};
\node[stage3, below=of postprocessing, xshift=1.5cm] (prepareState) {\textboldcolor{white}{prepare state for 
next step}};

\begin{scope}[every path/.style={-, thick, blue!20!black!80!white}]
    \draw (start.south) -- (preprocessing.north);
    \draw (start.south) -- (solver.north);
    \draw (start.south) -- (postprocessing.north);
\end{scope}
\begin{scope}[every path/.style={-, thick, blue!35!black!65!white}]
    \draw (preprocessing.south) -- (processInput.north);
    \draw (preprocessing.south) -- (encodeInput.north);
    \draw (solver.south) -- (customgates.north);
    \draw (solver.south) -- (subcircuits.north);
    \draw (postprocessing.south) -- (saveResults.north);
    \draw (postprocessing.south) -- (prepareState.north);
\end{scope}
\begin{scope}[every path/.style={-, thick, blue!50!black!50!white}]
    \draw (subcircuits.south) -- (collision.north);
    \draw (subcircuits.south) -- (propagation.north);
    \draw (subcircuits.south) -- (macroCalc.north);
    
\end{scope}

\begin{scope}[every path/.style={->,ultra thick, blue!50!black!50!white}]
    \draw[dotted] (processInput.south) -- (-6.525,-5.55) ;
    \draw[dotted] (encodeInput.south) -- (-4.025,-5.55) ;
    \draw[dotted] (customgates.south) -- (-1.25,-5.55) ;
    \draw[dotted] (saveResults.south) -- (3.75,-5.55) ;
    \draw[dotted] (prepareState.south) -- (6.25,-5.55) ;
    
\end{scope}

\end{tikzpicture}
\caption{Illustration of progressive decomposition of the quantum algorithm into smaller implementable modules - specifically for QLBM decomposing into modules such as Custom gates and Subcircuits shown.}
\Description{Figure 5 illustrates the hierarchical structure of a quantum algorithm, specifically for the Quantum Lattice Boltzmann Method (QLBM). The figure is divided into three main stages:
1. Pre-processing: This stage involves processing and encoding input data.
2. Solver: This stage includes custom gates and subcircuits, which are further broken down into collision, propagation, and macroscopic variable calculation.
3. Post-processing: This stage involves extracting results and preparing the state for the next step.
Each stage is progressively decomposed into smaller, more detailed components, promoting modularization and easier management. The flowchart provides a clear hierarchical view from high-level algorithm structure down to specific computational tasks.}
\label{fig:flowchart-decomposition}
\end{figure}
The flowchart in Figure~\ref{fig:flowchart-decomposition} outlines the structure of a quantum algorithm, divided into three main stages: Pre-processing, Solver, and Post-processing. In the pre-processing stage, input data is processed and encoded. The solver stage includes custom gates and subcircuits, with subcircuits further broken down into collision, propagation, and macroscopic variable calculation. The post-processing stage involves saving results and preparing the state for the next step. Each level of the flowchart provides a clear hierarchical view, from high-level algorithm structure down to specific computational tasks.

The quantum algorithm can be modularized according to the flowchart, allowing for clear separation of tasks and easier management. This ensures each stage is independently developed and tested, facilitating parallel development. The hierarchical structure promotes scalability and adaptability, making it suitable for various quantum algorithms by customizing each module to fit specific requirements.

\subsubsection{Pre-processing}\label{preprocess}

The quantum lattice Boltzmann method requires that initial values for the advection-diffusion problem be provided in the form of an amplitude vector. These initial values can be arbitrary and may vary according to the specific problem. The Intel Quantum SDK does not include a built-in amplitude encoding algorithm, and this makes it necessary to either to build one with the SDK or use some external amplitude encoding routine. 
There are several state preparation algorithms documented in the literature, such as the ones referenced in \cite{Shende}, \cite{Optimal_state_preparation}, and \cite{zylberman2023efficient}. To study the hybrid time-stepping implementation on the full-state simulator, we use the Qiskit~\cite{Qiskit} state preparation class as an external routine, which is built upon the reverse iterative procedure proposed by \cite[Section 4]{Shende}.

We implement this step using the Python interface which allows the conversion from OpenQASM to C\texttt{++} source code compatible with the Intel Quantum SDK. The flowchart in Fig.~\ref{fig:Timestepping} illustrates the steps involved in encoding the initial state. Note that any other amplitude encoding routine could just as well be used in the place of the Qiskit algorithm; we have simply chosen it out of convenience.

\begin{algorithm}
    \begin{algorithmic}[1]
        \caption{Encoding}
        \Require initial state = [] , num-qubits, lattice-size
        \Ensure Qasm  file  with  Quantum  Circuit
        \State normalize initial state
        \State initialize quantum Circuit (qc) 
        \State (qc) use qiskit initialize function with specified num-qubits and initial state
        \State resulting qc passed to Qiskit transpile
        \State transpiled circuit converted and saved as OpenQASM file
    \end{algorithmic}
\end{algorithm}

\subsubsection{Solver}

This module gathers together the implementation of four algorithmic steps: encoding, collision, propagation, and the calculation of macroscopic variables. Section~\ref{preprocess} outlined the implementation of the encoding step. We then have three steps to implement inside the SDK, and we divide each step into subcircuits with different quantum kernels, allowing for flexibility with the circuit parameters. Additionally, multi-qubit custom gates are generated for these subcircuits. The generation of these subcircuits culminates in the main function which calls all the defined subcircuits, as outlined in Algorithm~\ref{mainfunc}.
The subcircuits are formed using the custom gates and the definition of custom gates is given below.
\begin{algorithm}
    \begin{algorithmic}[1]
        \caption{Solver: Main function}
        \label{mainfunc}
        \Require subcircuits
        \Ensure probabilities
        \State initialize the backend
        \State quantum kernel encoding \Comment{Calling the subcircuits}
        \State quantum kernel collision
        \State quantum kernel propagation
        \State quantum kernel macros
        \State get probabilities and amplitude and save the data
    \end{algorithmic}
\end{algorithm}

\paragraph{Initialize backend}

In the main function, the initial step involves setting up a backend to act as our quantum hardware. The Intel Quantum SDK provides a full-state simulator, which serves as our window to testing the algorithm in ideal setting. 

Once we have access to actual quantum hardware, we can set up and prepare the backend to work with that specific type of quantum hardware. This is really the only significant change we will need to make. The core structure, algorithms, and methods we are using and discussing can be consistently reused regardless of the backend. Essentially, adjusting to real quantum hardware mainly involves configuring the backend while keeping everything else intact.

\paragraph{Custom gates}\label{customgates}

The Intel Quantum SDK provides a set of basic logic gates, with the help of these logic gates the different multi-qubit gates can be derived. We use these custom gates as buildings block for the subcircuits.

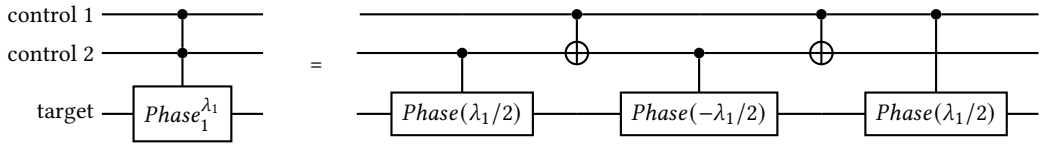
\begin{figure*}
    \centering
\begin{quantikz}[column sep=12pt, row sep=8pt, inner sep=2pt]
    \lstick{control 1} & \ctrl{1} & \qw & 
    \midstick[3,brackets=none]{=\phantom{00}} & 
    \qw & \ctrl{1} & \qw & \ctrl{1} & \ctrl{2} & \qw & \\
    \lstick{control 2} & \ctrl{1} & \qw & 
    & 
    \ctrl{1} & \targ{}  & \ctrl{1}  & \targ{} & \qw & \qw \\
    \lstick{target} & \gate{Phase_{1}^{\lambda_{1}}} & \qw & 
    &
    \gate{Phase(\lambda_{1} /2)} & \qw & \gate{Phase(-\lambda_{1} /2)} & \qw & \gate{Phase(\lambda_{1} /2)} & \qw  
\end{quantikz}
    \caption{Decomposition of two-qubit controlled phase gate : CCPHASE}
    \label{twoc phase}
    \Description{.}
\end{figure*}

In the quantum circuit given in Fig.~\ref{fig:Fullcircuit}, for the collision subcircuit there are two-qubit phase and two-qubit $X$ gates. The two-qubit $X$ gate or Toffoli is one of the logic gates in the set of given basic logic gates. The two-qubit controlled phase gate can be derived by using one- and two-qubit gates from the gate set. The decompositions illustrated for a two-qubit phase gate in Fig.~\ref{twoc phase} and for a five-qubit controlled $X$ gate in Fig.~\ref{five-qubit-not} are based on the general multi-qubit controlled gate decomposition explained in \cite[Section 4.3]{QuantumBible}. The pseudo-code for the two-qubit-controlled phase gate decomposition is given in Algorithm~\ref{twophasegate}. With this formulation we have created a basic building block custom gate named {\fontfamily{cmtt}\selectfont CCPHASE}.
\begin{algorithm}
    \begin{algorithmic}[1]
        \caption{Custom gate: CCPHASE -- two-qubit controlled phase gate}
        \label{twophasegate}
        \Require const int (control qubit1,control qubit2, target qubit, phase angle)
        \Ensure basic gate
        \Function {(quantum kernel) CCPHASE} {control 1, control 2, target qubit, angle}
            \State CPhase(qubit control 2, target,  angle / 2);
    	\State CX(qubit control 1, qubit control 2);
    	\State CPhase(qubit control 2, target,  - angle / 2);
    	\State CX(qubit control 1, qubit control 2);
    	\State CPhase(qubit control 1, target,  angle / 2);
        \EndFunction
    \end{algorithmic}
\end{algorithm}

Similarly, for the multi-qubit gates, there is the additional requirement of ancilla qubits as shown in the five-qubit example in Fig.~\ref{five-qubit-not}. The number of ancilla qubits required is one less than the number of control qubits. The pseudo-code for this particular five-qubit decomposition is given in Algorithm~\ref{multicnot}.
\begin{algorithm}
    \begin{algorithmic}[1]
        \caption{Custom gate: five-qubit controlled $X$ gate} \label{multicnot}
        \Require const int (control 1 - 5, target, phase angle)
        \Ensure basic gate
        \Function {(quantum kernel) MCX5} {control 1 - 5, target qubit, ancilla 1 - 4} \Comment{ 5 qubit controlled X gate}
            \State Toffoli(control 1, control 2, ancilla 1)
            \State Toffoli(control 3, ancilla 1, ancilla 2)
            \State Toffoli(control 4, ancilla 2, ancilla 3)
            \State Toffoli(control 5, ancilla 3, ancilla 4)
            \State CX(ancilla 4, target)   
            \State Toffoli(control 5, ancilla 3, ancilla 4)
            \State Toffoli(control 4, ancilla 2, ancilla 3)
            \State Toffoli(control 3, ancilla 1, ancilla 2)
            \State Toffoli(control 1, control 2, ancilla 1)
        \EndFunction
    \end{algorithmic}
\end{algorithm}
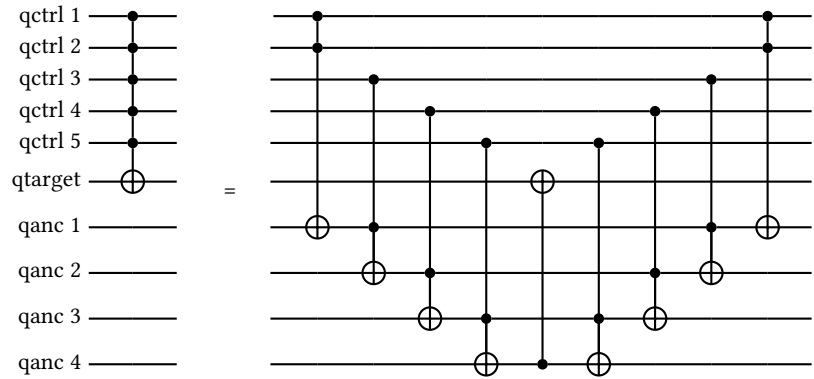
\begin{figure}[ht]
    \centering
\begin{quantikz}[column sep=12pt, row sep=8pt, inner sep=2pt]
    \lstick{qctrl 1} & \ctrl{1} & \qw & 
    \midstick[10,brackets=none]{=\phantom{00}} &
    \ctrl{1} & \qw & \qw & \qw & \qw & \qw & \qw & \qw & \ctrl{1} & \qw \\
    \lstick{qctrl 2} & \ctrl{1} & \qw &
    &
   \ctrl{5} & \qw & \qw & \qw & \qw & \qw & \qw & \qw & \ctrl{5} & \qw \\
    \lstick{qctrl 3} & \ctrl{1} & \qw &
    &
    \qw & \ctrl{5} & \qw & \qw & \qw & \qw & \qw & \ctrl{5} & \qw & \qw \\
    \lstick{qctrl 4} & \ctrl{1} & \qw &
    & 
    \qw & \qw & \ctrl{5} & \qw & \qw & \qw & \ctrl{5} & \qw & \qw & \qw \\
    \lstick{qctrl 5} & \ctrl{1} & \qw & 
    &
    \qw & \qw & \qw & \ctrl{5} & \qw & \ctrl{5} & \qw & \qw & \qw & \qw \\
    \lstick{qtarget} & \targ{} & \qw &
    &
    \qw & \qw & \qw  & \qw & \targ{}& \qw & \qw & \qw & \qw & \qw \\
    \lstick{qanc 1 } & \qw & \qw & 
    & 
    \targ{}& \ctrl{1} & \qw & \qw & \qw & \qw & \qw & \ctrl{1} & \targ{}& \qw \\
    \lstick{qanc 2 } & \qw & \qw & 
    &
    \qw & \targ{}& \ctrl{1} & \qw & \qw & \qw & \ctrl{1} & \targ{}& \qw & \qw \\
    \lstick{qanc 3 } & \qw & \qw & 
    &
    \qw & \qw & \targ{}& \ctrl{1} & \qw & \ctrl{1} & \targ{}& \qw & \qw & \qw \\ 
    \lstick{qanc 4 } & \qw & \qw & 
    & 
    \qw & \qw & \qw & \targ{}& \ctrl{-4} & \targ{}& \qw & \qw & \qw & \qw
\end{quantikz}
    \caption{Decomposition of five-qubit controlled $X$ gate (MCX5) using ancilla qubits}
    \label{five-qubit-not}
    \Description{.}
\end{figure}

\paragraph{Subcircuits}

In the Section~\ref{customgates} we explained the basic building blocks needed for constructing the subcircuits. This section elucidates the formation of different subcircuits mentioned in Section~\ref{steps} using custom gates and the basic gates provided by the SDK. The process of constructing the subcircuits is identical to that of custom gates, the only difference being that the subcircuits are one level higher and make use of the custom gates. It is important to note that the qubit parameters for the gates must be unambiguous, allowing the compiler to differentiate between quantum and classical instructions. Currently, this is a disadvantage when creating generalized quantum custom gates, as it requires precise qubit specification during the design.

\begin{algorithm}
    \begin{algorithmic}[1]
        \caption{Subcircuit: Collision}
        \label{code-collision}
        \Require const int (control qubit1,control qubit2, target qubit, phase angles 1-2)
        \Ensure collision subcircuit
        \Function {(quantum kernel) collision} {control 1, control 2, target qubit, angles 1-2}
            \State H(qubit control 1)
            \State X(qubit control 1)
            \State X(qubit control 2) \Comment{Changing the control value of qubit from 1 to 0 using X gate}\\
            
            \State CCPHASE(control 1, control 2, target qubit, angle 1)
            \State Toffoli(control 1, control 2, target qubit)
            \State CCPHASE(control 1, control 2, target qubit, angle 1)
            \State Toffoli(control 1, control 2, target qubit)\\

            \State X(qubit control 1) \\
            \State CCPHASE(control 1, control 2, target qubit, angle 2)
            \State Toffoli(control 1, control 2, target qubit)
            \State CCPHASE(control 1, control 2, target qubit, angle 2)
            \State Toffoli(control 1, control 2, target qubit)\\

            \State X(qubit control 1)
            \State X(qubit control 2)\\

            \State CCPHASE(control 1, control 2, target qubit, -angle 1)
            \State Toffoli(control 1, control 2, target qubit)
            \State CCPHASE(control 1, control 2, target qubit, -angle 1)
            \State Toffoli(control 1, control 2, target qubit)\\

            \State X(qubit control 1)\\
            
            \State CCPHASE(control 1, control 2, target qubit, -angle 2)
            \State Toffoli(control 1, control 2, target qubit)
            \State CCPHASE(control 1, control 2, target qubit, -angle 2)
            \State Toffoli(control 1, control 2, target qubit)\\

            \State H(qubit control 1)
        \EndFunction
    \end{algorithmic}
\end{algorithm}
The collision subcircuit requires the two-qubit controlled phase gate and a Toffoli with Hadamard gates on either side. The {\fontfamily{cmtt}\selectfont CCPHASE} custom gate is used to implement this subcircuit. 
The subcircuit also contains some controls with a value of zero instead of one. To convert the control value from one to zero, the $X$ gate is applied before and after the gate on the control qubit. The subcircuit is outlined in pseudo-code in Algorithm~\ref{code-collision}.

\begin{figure}[h]
    \centering
\begin{quantikz}[column sep=12pt, row sep=8pt, inner sep=2pt]
    \lstick{$q_0$} & \ctrl{1} & \ctrl{1} & \ctrl{1} & \targ{}& \qw  & \ctrl{1} & \qw \\
    \lstick{$q_1$} & \ctrl{1} & \ctrl{1} & \ctrl{1} & \qw & \targ1 & \targ2 & \qw  \\
     \lstick{$q_2$} & \ctrl{1} & \ctrl{1} & \targ{}& \qw & \qw & \qw & \qw \\
    \lstick{$q_3$} & \ctrl{1} & \targ{}& \qw & \qw & \qw & \qw & \qw \\
     \lstick{$q_4$} & \targ{}& \qw & \qw & \qw & \qw & \qw & \qw \\
    \lstick{$q_5$} & \octrl{-1} & \octrl{-2} & \octrl{-3} & \octrl{-5} & \octrl{-4} & \octrl{-4} & \qw \\
\end{quantikz}
    \caption{The right shift operator for a circuit of five working qubits. The last qubit controls on the direction of the shift}
    \label{fig:rshift}
    \Description{.}
\end{figure}
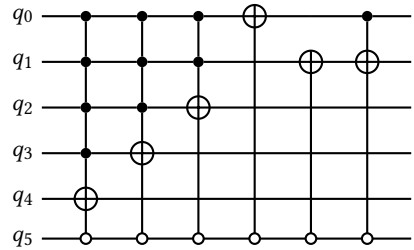

For the propagation subcircuit, further division into two modules (subcircuits) is possible based on the streaming direction, left and right. As an example we explain the construction of the right shift operator, shown in the case of five working qubits in Fig.~\ref{fig:rshift}. This right shift subcircuit consists of three, four, and five qubit-controlled $X$ gates. The Intel Quantum SDK enables us to create different custom gates for distinct numbers of control qubits in the general case.
Here, various custom gates with {\fontfamily{cmtt}\selectfont MCX5, MCX4,} and {\fontfamily{cmtt}\selectfont MCX3} are constructed following the steps provided in Section.~\ref{customgates}. These custom gate quantum kernels are invoked in the subcircuit of the right shift operator. The pseudocode in Algorithm~\ref{shiftcode} demonstrates this particular implementation.
\begin{algorithm}
    \begin{algorithmic}[1]
        \caption{Subcircuit: five-qubit right shift}
        \label{shiftcode}
        \Require const int (control qubits, target qubits)
        \Ensure right shift subcircuit
        \Function {(quantum kernel) right\textunderscore{}shift} {control qubits, target qubits}
            \State X(qubit control)
            \State MCX5(control qubits, target)
            \State MCX4(control qubits, target)
            \State MCX3(control qubits, target)
            \State CX(control, target)
            \State CX(control, target)
            \State Toffoli(control qubits, target)    
            \State X(qubit control)
        \EndFunction
    \end{algorithmic}
\end{algorithm}

Constructing the subcircuit and custom gates help us divide the complex algorithm in smaller modules. It helps in writing generalized and expandable programs on a higher level, giving an optimized way for writing quantum circuits. Coupling these subcircuits together as illustrated in Algorithm~\ref{mainfunc}, we complete the whole algorithm for the QLBM example.

\subsubsection{Post-processing}

Post-processing module consist of extracting the required probabilities from the full-state simulator and deriving the macroscopic concentration values. 
The process to derive the concentration from the probabilities is given in \cite[Section 3.1]{ADE}, in particular the renormalization of the state amplitudes. These values are then fed back to the algorithm as the initial state for the next time step, until the desired time evolution is completed.

\subsection{Expanding to 2D advection-diffusion and Navier-Stokes equations}
\label{sec:to-navier-stokes}

The implementation of the 1D quantum algorithm involves formulating custom gates that can be generalized, resulting in a set of generic multi-qubit gates. This approach allows for the reuse of these custom gates in other algorithms, and new custom gates can similarly be formulated to suit other quantum circuits. In this spirit, we briefly outline an expansion to the 2D advection-diffusion~\cite{ADE} and Navier-Stokes~\cite{NavierQLBM} algorithms. These algorithms share several similarities with the 1D advection-diffusion, which simplifies their construction. 

For the 2D advection-diffusion, which follows a D2Q5 lattice arrangement, the difference is mainly in the collision subcircuit with five multi-qubit controlled operators, one for each velocity. The process of creating these operators is similar to 1D: formulate custom gates (to be specific, a three-qubit controlled phase gate) and utilize them to derive the subcircuit for the 2D collision. Similarly for the propagation step, the circuit is divided into four smaller modules according to the directions of the necessary four shifts. Combining them together similarly to the 1D case forms the propagation subcircuit. Many custom gates from the 1D implementation can be reused but some new gates are also needed.

The implementation of the Navier-Stokes QLBM algorithm involves solving two lattice Boltzmann equations simultaneously as explained in~\cite{NavierQLBM}. The collision subcircuit is completely different from the one used in 2D advection-diffusion equation, while the propagation step is carried out twice respectively for the two governing equations. The primary challenge lies in the collision subcircuit, which requires classical calculation of the diagonal elements of the non-unitary matrix and further encoding the matrix using the linear combination of operators approach. The efficient hybrid quantum-classical design is particularly valuable here. The remaining parts of the algorithm can be implemented using the modular approach (custom gates and subcircuits) in the vein of the previous discussion.

\subsection{Numerical validation of the circuits}

In this section we implemented the quantum algorithm for 1D advection-diffusion in the Intel Quantum SDK while demonstrating the use of classical programming principles in quantum programming.  
Further, for a simple numerical validation of the soundness of the implementation as a physics model, we compare the results after each step with the known analytical solution and the classical lattice Boltzmann method.

The example problem considered is a transport in a channel having a fluid concentration $C(x,t)$. A Gauss-like distribution sets the initial concentration $C(x,0)$ for the problem. The setup is the following:
\begin{enumerate}
    \item A periodic boundary condition.
    \item Number of lattice sites is a power of 2, the size is chosen so that the distributions do not overlap. We use a minimum of $2^4 = 16$ lattice sites.
    \item For the illustration here, the number of lattice sites is $32$ and the initial concentration is given as $C(5,0) = C(7,0) = 0.5$ and $C(6,0) = 1.0$.
\end{enumerate}
 
\begin{figure}[ht]
    \centering
    \begin{subfigure}[b]{0.495\textwidth}
        \centering
        \includegraphics[
        width=\textwidth]{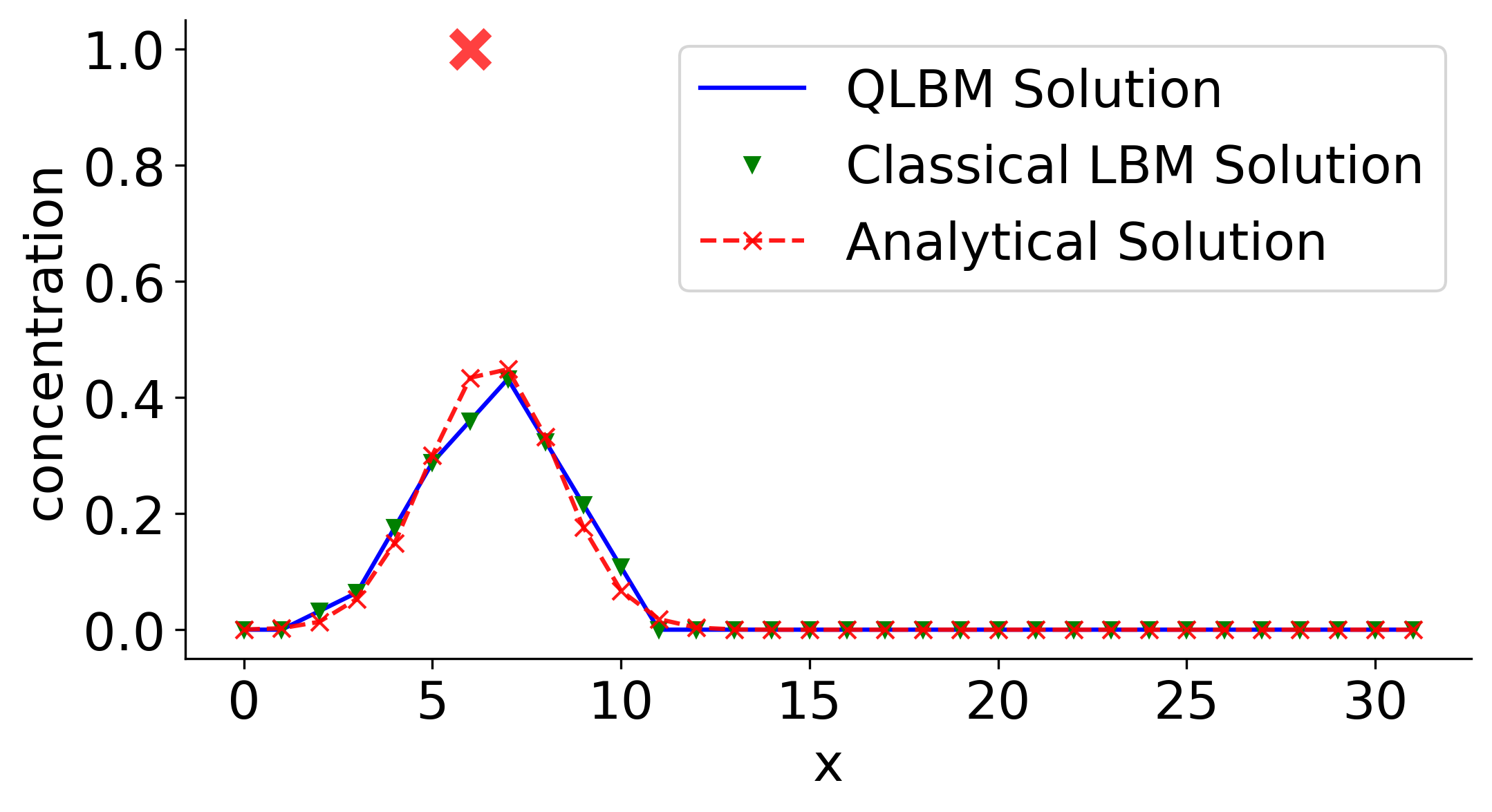}
        \caption{Step 2}
    \end{subfigure}
    \hfill
    \begin{subfigure}[b]{0.495\textwidth}
        \centering
        \includegraphics[
        width=\textwidth ]{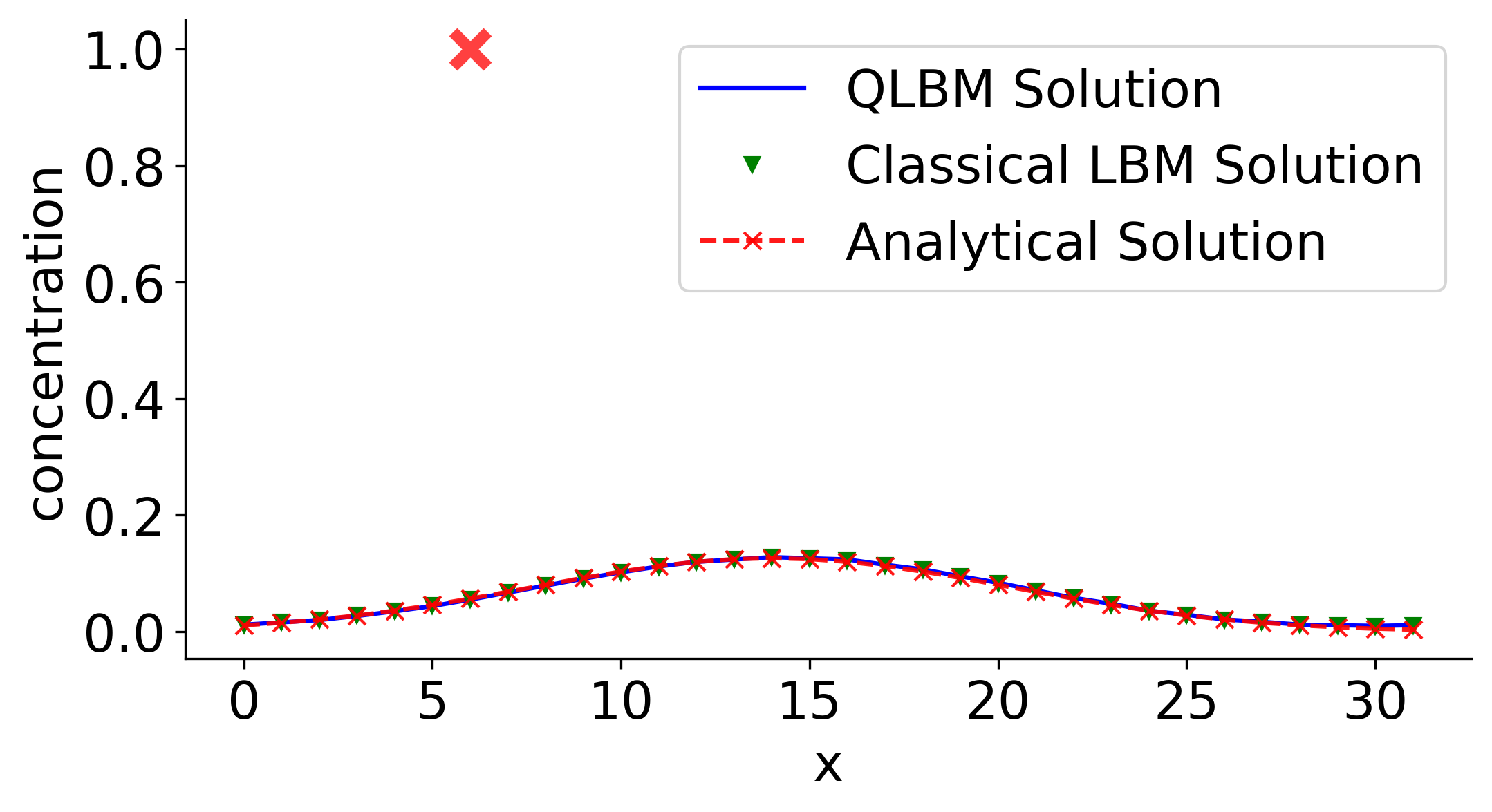}
        \caption{Step 40}
    \end{subfigure}
    \Description{.}
    \caption{Comparison between the analytical (denoted by a dotted red line with cross mark), QLBM (denoted by a blue line) and the classical lattice Boltzmann method (denoted by green triangle) results for time steps $2$ and $40$. The standalone red cross at the top denotes the concentration $C(6,0) = 1.0$ as the initial condition.}
        \label{results}
\end{figure}

In Fig.~\ref{results} we visualize the concentrations extracted after time steps $2$ and $40$. 
The initial concentration is a triangle-shape with maximum concentration at a lattice site $x=6$ (denoted with a red cross), gradually decreasing towards the edge of the domain. The results show some difference with the analytical solution in the first few time steps. This is not produced by the quantum algorithm itself but is a characteristic of the Boltzmann discretization and the low resolution of the computational grid. To some extent the accuracy of the model could be increased with a finer grid providing a higher resolution and, hence, requiring a greater number of qubits for the circuit implementation. We compared these results with the classical lattice Boltzmann method and they agree within a numerical tolerance of $10^{-12}$ which is ignorable, which confirms that the discrepancies are due to the inherent limitations of the discretization and grid resolution.

\section{Discussion: Using classical techniques on quantum SDKs}

The Intel Quantum SDK differs from other quantum software development kits like Qiskit\cite{Qiskit}, PennyLane \cite{pennylane}, and the \textit{Munich Quantum Toolkit} (MQT)~\cite{mqtbench} in several key aspects, including its target hardware, programming approach, and integration with classical computing resources. In essence, the Intel Quantum SDK is distinct in its integration with the Intel classical computing resources and its focus on high-performance and hybrid quantum-classical computing, leveraging \texttt{C++} for development. In contrast, Qiskit, PennyLane, and MQT are more flexible in terms of hardware support and focus on accessibility and integration with programming languages like Python, each with its own specialized use cases and target communities. The classical programming principles outlined can be applied to these quantum SDKs with some adjustments to fit the specific features and capabilities. Here is a brief overview of the aforementioned SDKs from this perspective:
\begin{enumerate}
    \item Modularization: 
    All Qiskit, Pennylane and Munich Quantum Toolkit support modular programming, allowing developers to create reusable components. This principle can be applied by structuring quantum programs into functions or classes that can be easily reused and maintained.
    \item Hybrid design: These SDKs also support hybrid quantum-classical computing, enabling the integration of quantum algorithms with classical pre- and post-processing. Developers can design algorithms that leverage the strengths of both quantum and classical computing.
    \item Custom gates: While the specific method of creating custom gates may vary between SDKs, the concept remains the same. Developers can define new gates as combinations of existing gates to simplify complex circuit designs.
    \item Iterative development: The iterative approach to developing and refining quantum algorithms is universally applicable. Developers can iteratively test and optimize their quantum circuits in any SDK to achieve the desired outcomes.
\end{enumerate}
Overall, the principles of classical programming provide a solid foundation for developing quantum algorithms in any quantum SDK, promoting code reusability, maintainability, and efficiency. The key is to adapt these principles to the unique features and constraints of the chosen SDK. In this work, the Intel Quantum SDK has been utilized due to its emphasis on adapting such features to the \texttt{C++} programming language, particularly crafted to support quantum operations to bridge the gap between traditional programming and the unique requirements of quantum computing~\cite{IntroIQSDK-IEEE-2023}. In our experience, this integration of classical programming paradigms with quantum functionalities creates a user-friendly and efficient hybrid quantum-classical programming experience, making it accessible to a broad range of users. However, as mentioned above, similar efforts are commonplace in the quantum computing community, and we do not aim to (nor can we) place these efforts in any particular order of preference based on this work.

\section{Summary}

In this paper, we recreated the quantum algorithm for solving the advection-diffusion equation using the lattice Boltzmann method while explaining the principles used to generalize and expand the quantum circuits with the Intel Quantum SDK.
It was shown how classical programming techniques, such as iterative development cycles and modular design methods, can be leveraged in quantum programming. The top-down modular design approach helps break down complex quantum algorithms into subcircuits, while the creation of custom gates aids in implementing multi-qubit controlled gates. The creation of generalized circuits in terms of number of qubits and the problem variables benefits from this approach. 
In particular, these techniques enable one to maintain and reuse quantum kernels in the Intel Quantum SDK. 
This is an example of how the utilization of classical software development methodologies helps to achieve robust and beneficial quantum software code.

\section*{Acknowledgment}

We would especially like to thank and appreciate the entire Intel family for their key support. Among these are Prof.~Dr.~Anne Matsuura and Dr.~Kevin Rasch who were consistently guiding and encouraging us throughout the process. Lastly, T.S. would like to appreciate all his colleagues and teachers for their academic and moral support throughout the journey. This research was partially supported by the Business Finland project 9820/31/2022 Quantum-Native Multiphysics.


\printbibliography
\section*{Appendix}
\appendix

\section{Linear Combination of Unitary matrix approach}
The linear combination of unitaries (LCU) approach is a technique used in quantum computing to efficiently simulate quantum circuits and algorithms. This approach involves representing a quantum operation as a linear combination of elementary unitary operations. Unitary operations are fundamental to quantum mechanics and represent reversible transformations on quantum states.

Mathematically, this can be expressed as follows:

Suppose we have a quantum operation $U$ that we want to represent as a linear combination of unitaries $U_1, U_2, \ldots, U_n$. Then, we can write:

\[ U = \sum_{i=1}^{n} c_i U_i \]

where $c_i$ are complex coefficients representing the proportions of each unitary operation $U_i$ in the combination.

Here is a more detailed breakdown:

\begin{itemize}
  \item $U$: The target quantum operation we want to represent.
  \item $U_1, U_2, \ldots, U_n$: Elementary unitary operations or a set of predefined unitaries.
  \item $c_1, c_2, \ldots, c_n$: Complex coefficients determining the contribution of each unitary operation in the linear combination.
\end{itemize}

In practice, determining the coefficients $c_i$ can involve various techniques such as optimization algorithms, theoretical analysis, or experimental measurements.

LCU is particularly useful in quantum computing for efficiently simulating quantum circuits and algorithms, as it allows for the decomposition of complex quantum operations into simpler unitary operations, making simulations more tractable and scalable.

Let's consider the example of a single-qubit quantum operation, the Hadamard gate (\(H\) gate). The Hadamard gate is a fundamental gate in quantum computing that creates superposition. It can be represented by the following matrix:

\[
H = \frac{1}{\sqrt{2}} \begin{bmatrix}
1 & 1 \\
1 & -1 \\
\end{bmatrix}
\]

Now, let's decompose the Hadamard gate into a linear combination of elementary unitary operations. We can represent the Hadamard gate as a linear combination of the Pauli matrices, which are elementary unitary operations. The Pauli matrices are denoted as:

\[
\sigma_x = \begin{bmatrix}
0 & 1 \\
1 & 0 \\
\end{bmatrix}
\]

\[
\sigma_y = \begin{bmatrix}
0 & -i \\
i & 0 \\
\end{bmatrix}
\]

\[
\sigma_z = \begin{bmatrix}
1 & 0 \\
0 & -1 \\
\end{bmatrix}
\]

Now, we can express the Hadamard gate (\(H\)) as a linear combination of the Pauli matrices:

\[
H = \frac{1}{\sqrt{2}}\sigma_x + \frac{1}{\sqrt{2}}\sigma_z
\]

This representation shows that the Hadamard gate can be decomposed into a linear combination of the Pauli matrices \(\sigma_x\) and \(\sigma_z\).

\end{document}